\documentclass[pra,twocolumn,
superscriptaddress,showpacs]{revtex4}
\usepackage{amsmath}
\usepackage{amsfonts}
\usepackage{graphicx}
\usepackage{epsfig}
\errorstopmode

\begin{document}

\title{Efficient sharing of a continuous-variable quantum secret}

\author{Tom\'a\v{s} Tyc}
\affiliation{Department of Physics and Centre for Advanced Computing --
Algorithms and Cryptography, \\
Macquarie University, Sydney, New South Wales 2109, Australia}
\affiliation{Institute of Theoretical Physics, Masaryk
                 University, Kotl\'a\v rsk\'a 2, 61137 Brno, Czech Republic}
\author{David J. Rowe}
\affiliation{Department of Physics and Centre for Advanced Computing --
Algorithms and Cryptography, \\
Macquarie University, Sydney, New South Wales 2109, Australia}
\affiliation{Department of Physics, University of
 Toronto, Toronto, Ontario M5S 1A7, Canada}
\author{Barry C.~Sanders}
\affiliation{Department of Physics and Centre for Advanced Computing --
Algorithms and Cryptography, \\
Macquarie University, Sydney, New South Wales 2109, Australia} 

\date{8 January 2003}

\begin{abstract}
We propose an efficient scheme for sharing a continuous variable
quantum secret using passive optical interferometry and squeezers:
this efficiency is achieved by showing that a maximum of two squeezers
is required to replicate the secret state, and we obtain the cheapest
configuration in terms of total squeezing cost. Squeezing is a cost for
the dealer of the secret as well as for the receivers, and we quantify
limitations to the fidelity of the replicated secret state in terms of
the squeezing employed by the dealer.
\end{abstract}
\maketitle

\section{Introduction} 
Secret sharing (SS) is an important cryptosystem protocol for dealing
secret information to a set of players, not all of whom can be trusted
\cite{Sha79}. The encoded secret can only be replicated (or,
equivalently, reconstructed~\cite{tomography}) if certain subsets of
players collaborate, and these subsets are referred to as the access
structure. The remaining subsets comprise the adversary structure, and
the protocol denies the adversary structure any information about the
secret.  The underpinning scheme for arbitrary SS is $(k,n)$-threshold
SS, which involves $n$ players, and any subset of~$k$ players
constitutes a valid set in the access structure; other SS
schemes can be constructed via threshold SS, for example
by distributing unequal numbers of shares to players.  Although
quantum secret sharing (QSS) was first introduced as a method to
transmit classical information in a hostile environment with
quantum-enhanced security \cite{Hil99}, QSS was subsequently
established \cite{Cle99} as a quantum analogue to Shamir's secret
sharing described above and we use the term QSS to refer to the latter
approach.  QSS provides a valuable protocol in quantum communication
but also is important as an error correction scheme \cite{Cle99}.

Here we are concerned with continuous-variable (CV) QSS~\cite{Tyc02}.
Quantum information protocols and tasks are now studied both as
discrete-variable, qubit-based (or qudit-based) protocols and
tasks~\cite{Nie00} and as CV realizations~\cite{Bra02}.  CV quantum
information protocols are generally realized in optical systems and
exploit advanced quantum optics tools, such as the generation of
squeezed light \cite{Lou87} and ability to count single photons
\cite{Kim99,Bar02}, as well as the low rate of decoherence for optical
systems. The recent demonstration of CV unconditional quantum
teleportation \cite{Fur98} is an excellent example of the capabilities
of CV quantum information processes in optical systems. Moreover, the
technology for this CV quantum teleportation is not very different
from the techniques required for CV QSS.

The original proposal for CV QSS \cite{Tyc02} established a general
method for CV QSS, and for $(k,n)$-threshold schemes in particular,
using interferometry involving passive optical elements (mirrors, beam
splitters and phase shifters), active elements (squeezers) and
homodyne detectors. A $(2,3)$-threshold scheme was proposed involving
a single squeezer, thereby suggesting an experiment that is within the
reach of current technology~\cite{Lan02}.  The original proposal of
how to perform the general $(k,n)$ scheme was complicated, though, by
the need for an increasing number of squeezers in the interferometer.
A practical realization of threshold-QSS would need to minimize the
number of optical squeezers as the number of players increases.

Here we establish that, for any number of players $n$ and any
threshold level $k$ for the number of collaborators to be in the
access structure, the total number of squeezers needed by the
collaborating players does not exceed two. This remarkable result
informs us that at most two squeezers are required for an arbitrary
number of players $n$.  In particular, to replicate the secret state,
the collaborating players require access to an interferometer with $k$
channels but only two active components (i.e., squeezers).  This
analysis also allows us to determine the total amount of squeezing
required in a two-squeezer threshold QSS protocol: the analysis is
important because the degree of squeezing required for the protocol
can be regarded as an effective cost for the procedure \cite{Bra99}.

The second major concern of this paper is the extent to which it is possible
to achieve the goals of the CV QSS protocol with finite physical
resources. For the protocol to work perfectly, the dealer needs access
to ancillary states prepared with infinite squeezing; as this is not
physically possible, we analyze the effects of finite squeezing, which
imposes limitations on the fidelity of the replicated secret state.

The paper is organized as follows: in Section~\ref{threshold} we
summarize the CV QSS protocol for threshold schemes. In
Section~\ref{weakening} we describe efficient replication of the
secret state, which requires the minimal number of squeezing elements and
minimal overall squeezing.
The total amount of squeezing is discussed in Section~\ref{total} and
we conclude in Section~\ref{conclusion}.

\section{Threshold QSS with Finite resources}
\label{threshold} 

\begin{figure}[ht]
\epsfig{file=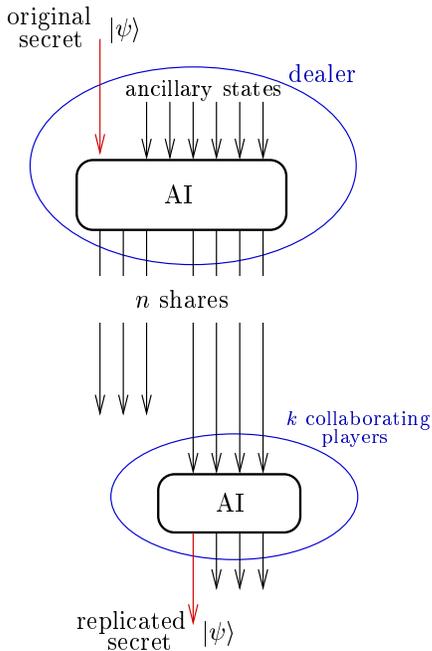, height=3.4 in}  
\caption{The optical $(k=4,n=7)$ QSS threshold scheme: the dealer
  encodes the secret via an active interferometer (AI) by mixing it
  with $n-1$ ancillary states, transmits the resulting $n$ shares to
  the players, and any $k$ players employ a second interferometer to
  replicate the secret state.  The interferometers are active, meaning
  that they employ both passive optical devices and energy-consuming
  squeezers.  }
\label{scheme}\end{figure}

The optical $(k,n)$ threshold scheme is sketched in Fig.~\ref{scheme}.
A dealer holds a pure secret state $|\psi\rangle$ realized in a single
mode of the electromagnetic field and encodes the secret as an
$n$-mode entangled state $|\Psi\rangle$ by mixing with $n-1$ ancillary
modes in an $n$-channel active interferometer, where the term active
refers to one-- or two--mode squeezers~\cite{Sch85}.  The dealer then
sends one output, or ``share'', to each of the players, and at least
$k$ players must combine their shares in an active interferometer to
replicate the secret state.  However, the no-cloning
theorem~\cite{Woo82} requires that no threshold scheme exists for
$n\ge 2k$~\cite{Cle99}.  Also any threshold scheme with $n<2k-1$ can
be obtained from the $(k,2k-1)$ scheme by discarding $2k-1-n$ shares.
Therefore, we concentrate on the $(k,2k-1)$ threshold scheme.

\subsection{Entanglement of the secret state}

The secret is a state $|\psi\rangle\in \mathbb{H}^{(1)}\sim
\mathcal{L}^2(\mathbb{R})$ with wave function $\psi(x)=\langle x|\psi\rangle$.
Let $\mathbb{H}^{(n)}$ be the tensor product of $n=2k-1$ copies of
$\mathbb{H}^{(1)}$, one copy of which is owned by each player. 
The idea is to entangle the information among states of
$\mathbb{H}^{(n)}$ such that any $k$ players can cooperate to untangle
the information but any smaller number is unable to do so.

The Hilbert space $\mathbb{H}^{(n)}$ is the space
$\mathcal{L}^2(\mathbb{R}^n)$ of square integrable wave functions on
$\mathbb{R}^n$. 
Thus, if  ${\mathbb F}^{n}$ denotes the linear space of coordinate
functions for $\mathbb{R}^n$, then choosing a system of Euclidean
coordinates $(x_1, \ldots, x_n)$ for any vector ${\bf x} \in
\mathbb{R}^n$ is equivalent to picking an orthonormal basis $(f_1, \ldots
, f_n)$ for ${\mathbb F}^{n}$ such that
\begin{equation}
 f_i ({\bf x}) = x_i \,.
\end{equation}
We denote the inner product of these coordinate
functions by $f_i \cdot f_j = \delta_{ij}$.

Suppose the dealer starts with an unentangled tensor product 
\begin{equation}
|\Psi\rangle = |\psi\rangle \otimes \underbrace{|\varphi_a\rangle
   \otimes \cdots\otimes|\varphi_a\rangle}_{k-1} \otimes 
 \underbrace{|\varphi_{1/a}\rangle \otimes \cdots 
\otimes|\varphi_{1/a}\rangle}_{k-1} \,,
\end{equation}
of the secret
state $|\psi\rangle$, with $k-1$ copies of a state $|\varphi_a\rangle$ and
$k-1$ copies of a state $|\varphi_{1/a}\rangle$, where  
\begin{equation}
\varphi_a(x) = \langle x |\varphi_a\rangle =
(\pi a^2)^{-1/4}\, e^{-x^2/2a^2}  \,.
\label{squeezed}\end{equation}
Write this state 
\begin{equation}
|\Psi\rangle = \int {\rm d}x^n\,  \Psi({\bf x})
\,|x_1\rangle \otimes \cdots \otimes|x_n\rangle \,,
\end{equation}
where
\begin{equation}
\Psi({\bf x}) = \psi(x_1) \prod_{i=2}^k \varphi_a(x_i) \prod_{i=k+1}^n
\varphi_{1/a}(x_i)
 \,.
\end{equation}

The dealer then entangles the secret state by a linear
canonical point transformation 
\begin{equation}
 f_i \to g_i =\sum_j g_{ij} f_j \,, \label{eq:entangle}
\end{equation}
in which the orthogonal (Euclidean) coordinate functions $\{f_i \}$ are
replaced by a general linear system $\{g_i \}$ for which
$g_i({\bf x}) = \sum_j g_{ij} f_j({\bf x}) = \sum_j g_{ij}x_j$.
The corresponding unitary transformation of $\mathbb{H}^{(n)}$ then
maps the state $|\Psi\rangle$ to
\begin{equation}
|\Psi_g\rangle = |\det g|^{1/2} \int {\rm d}x^n\,  \Psi({\bf x})
\,|g_1({\bf x}) \rangle \otimes \cdots \otimes|g_n({\bf x})\rangle \,.
\label{psi_g}\end{equation}
For it to be possible for any subset of $k$ players
to reconstruct the secret state, certain conditions  must be respected.
These conditions become apparent when we consider the replication
algorithm.

\subsection{The replication algorithm}

In replicating the secret state, it is convenient to identify three
subspaces of coordinates; i.e., express ${\mathbb F}^{n}$ as a direct sum
of three mutually orthogonal subspaces
\begin{equation}
{\mathbb F}^{n} = \mathbb{X} \oplus \mathbb{Y} \oplus \mathbb{Z}\,,
\label{eq:FXYZ}\end{equation}
where $\mathbb{X}$ is the one-dimensional space spanned by $f_1$, 
and $\mathbb{Y}$ and $\mathbb{Z}$ are the $(k-1)$-dimensional spaces
spanned, respectively, by $\{ f_2,\ldots , f_k\}$ and
$\{ f_{k+1}, \ldots , f_n\}$.
Thus, we relabel the $\{ x_i\}$ coordinates
as $(x,y_i,z_i)$ coordinates with
\begin{eqnarray}
 &x = x_1, &\nonumber\\ 
&y_i = x_{i-1}, \quad
z_i = x_{k+i}, 
 \quad i=2,\ldots k \,.
\end{eqnarray}
The wave function $\Psi$ is then
\begin{equation}
\Psi({\bf x}) = \psi(x) \prod_{i=1}^{k-1} \varphi_a(y_i) 
\varphi_{1/a}(z_i) \,.
\end{equation}

It will be understood in the following that all $n$ players know the
encoding transformation in which $f_i \to g_i$. We then impose the
requirement that this transformation is such that the components of
any $k$ basis vectors of the set $\{ g_i\}$ that lie in the subspace
$\mathbb{X} \oplus \mathbb{Y}\subset {\mathbb F}^{n}$ are linearly
independent and hence span this subspace.

Without loss of generality, we may suppose that the first $k$
players form the collaborating set. 
These players are able to make any transformation of the states in the
subset of Hilbert spaces accessible to them.
However, we will restrict the transformations they can make to those
corresponding to general linear coordinate transformations, as defined
above.  Let us suppose they make the transformation 
\begin{equation}
g_i \to \xi_i = \sum_j \xi_{ij}f_j
\end{equation}
with the understanding that $\xi_i = g_i$ for all $i> k$.

The orthogonal decomposition of ${\mathbb F}^{n}$ given by
Eq.~(\ref{eq:FXYZ}) now defines a corresponding decomposition of every
$\xi_i$ vector as a sum of three mutually orthogonal vectors
\begin{equation}
  \xi_i = \alpha_i + \beta_i  + \gamma_i \,.
\label{xi_i}\end{equation}
Equivalently, we can write 
\begin{equation}
  \xi_i ({\bf x}) = \alpha_i x + \sum_j \beta_{ij}y_j +\sum_j
\gamma_{ij}z_j \,.
\end{equation}

We now claim that a transformation $g_i \to \xi_i$ which is such
that
\begin{eqnarray}
 &\alpha_1 = 1, \quad \beta_1=0 ,&\nonumber\\
&\alpha_{i+1} = \alpha_{k+i} , \quad \beta_{i+1} =
\beta_{k+i} , \quad i = 1,\ldots, k-1\,,  & \label{eq:xi}
\end{eqnarray}
replicates the secret for sufficiently large values of
the parameter $a$.
We demonstrate this
result explicitly for the simple case in which $k=2$ and $n=3$.

For the $k=2$, $n=3$ case $(\xi_1,\xi_2,\xi_3)$ will have expansions of
the  form
\begin{eqnarray}
 &\xi_1({\bf x}) = x + \gamma_1 z, & \\
&\xi_2({\bf x})  = \alpha x +\beta y + \gamma_2 z, & \\
&\xi_3({\bf x})  = \alpha x +\beta y + \gamma_3 z, &
\end{eqnarray}
and $|\Psi_\xi\rangle$ will be given by 
\begin{eqnarray}
 |\Psi_\xi\rangle &=& \frac{|\beta(\gamma_2-\gamma_3)|^{1/2}}{\pi^{1/2}}
\int
\psi(x) 
\exp \big[ - \frac{1}{2a^2} y^2 - \frac{a^2}{2} z^2\big]\nonumber\\
& \times& |x + \gamma_1 z\rangle \otimes |\alpha x +\beta y + \gamma_2
z\rangle\otimes |\alpha x +\beta y + \gamma_3 z\rangle \nonumber\\
&\times& {\rm d}x\, {\rm d}y\, {\rm d}z\,. 
\end{eqnarray}
By a change of the variable $x$ to $x-\gamma_1 z$, we then have
\begin{eqnarray}
 |\Psi_\xi\rangle &=&
\frac{|\beta(\gamma_2-\gamma_3)|^{1/2}}{\pi^{1/2}} \int
\psi(x-\gamma_1z) \nonumber\\
&&\times \exp \big[ - \frac{1}{2a^2} y^2 - \frac{a^2}{2}
z^2\big]\nonumber\\ 
&& \times |x \rangle \otimes |\alpha x +\beta y +
\gamma'_2 z\rangle\otimes |\alpha x +\beta y + \gamma'_3
z\rangle\nonumber\\ &&\times {\rm d}x\, {\rm d}y\, {\rm d}z\,. 
\end{eqnarray}
Now observe that if $a$ is sufficiently
large that
$\psi(x-\gamma_1z)\approx
\psi(x)$ for all values of $z$ for which $\exp[-a^2z^2/2]$ is
non-negligible, then
\begin{eqnarray}
\psi(x-\gamma_1z) \exp \big[ - \frac{a^2}{2} z^2\big]\approx
\psi(x) \exp \big[ - \frac{a^2}{2} z^2\big] \,.
\end{eqnarray}
Moreover, this approximation becomes precise to any desired level of
accuracy for sufficiently large values of $a$.
By a second change of variables,
\begin{equation}
x\to x , \quad \beta y \to \beta y - \alpha x \,,
\end{equation}
 we also have
\begin{eqnarray}
 |\Psi_\xi\rangle &=&
\frac{|\beta(\gamma_2-\gamma_3)|^{1/2}}{\pi^{1/2}} \int
\psi(x) \nonumber\\
&\times& \exp \big[ - \frac{1}{2a^2} (y-\frac{\alpha}{\beta} x)^2 -
\frac{a^2}{2} z^2\big]\nonumber\\ & \times& |x \rangle \otimes |\beta y +
\gamma'_2 z\rangle\otimes |\beta y + \gamma'_3 z\rangle
\,{\rm d}x\, {\rm d}y\, {\rm d}z\,. 
\end{eqnarray}
Now for $a$  sufficiently large that
$\exp \big[ - \frac{1}{2a^2} (y-\frac{\alpha}{\beta} x)^2\big]
\approx \exp \big[ - \frac{1}{2a^2} y^2\big]$ for all values of $x$ for
which $\psi(x)$ is non-negligible, we have
\begin{eqnarray}
 |\Psi_\xi\rangle &\approx &
\frac{|\beta(\gamma_2-\gamma_3)|^{1/2}}{\pi^{1/2}} \int \psi(x) 
\exp \big[ - \frac{1}{2a^2} y^2 - \frac{a^2}{2}
z^2\big]\nonumber\\ & \times& |x \rangle \otimes |\beta y +
\gamma'_2 z\rangle\otimes |\beta y + \gamma'_3 z\rangle\,{\rm d}x\,
       {\rm d}y\, {\rm d}z
\nonumber\\
&=& |\psi\rangle\otimes |\Phi\rangle \,, 
\end{eqnarray}
where $|\Phi\rangle$ is the entangled state
\begin{eqnarray}
|\Phi\rangle &=& \frac{|\beta(\gamma_2-\gamma_3)|^{1/2}}{\pi^{1/2}} \int 
\exp \big[ - \frac{1}{2a^2} y^2 - \frac{a^2}{2}
z^2\big]\nonumber\\ & \times& |\beta y +
\gamma'_2 z\rangle\otimes |\beta y + \gamma'_3 z\rangle\, {\rm d}y\, {\rm d}z
\end{eqnarray}

The generalization of the proof to larger values of $k$ is
straightforward.

\subsection{Fidelity of the secret sharing scheme}

As we have seen, the CV QSS scheme works perfectly only for
$a\rightarrow\infty$ in Eq.~(\ref{squeezed}). In this case the dealer
has infinitely squeezed ancillary states with which to entangle the
secret state $|\psi\rangle$.  The situation is similar to CV quantum
teleportation \cite{Vai94}, where an ideal EPR pair (which is a
two-mode infinitely squeezed vacuum) is required for the protocol to
work perfectly.  However, with some loss of fidelity the scheme can be
adapted to a realistic, finite-squeezing situation \cite{Bra98}.  In
CV QSS, finite squeezing implies that the secret state can only be
approximately replicated because there is entanglement between the
secret state and the shares in both the access structure and the
adversary structure, which limits the fidelity of the replicated state
with respect to the original secret state. Also entanglement with the
adversarial shares allows some information about the secret state to
escape. These compromises to CV QSS are reduced by increasing the
degree of squeezing.

A detailed analysis reveals that the reduced density operator
$\hat\rho'$ of the replicated secret is related to the original
density operator $\hat\rho=|\psi\rangle\langle\psi|$ by
\begin{eqnarray} \nonumber
 \rho'(x,x')&\equiv&\langle x|\hat\rho'|x'\rangle=\frac{a}{\sqrt{\pi}\,v}\,
  \exp\left[-\frac{u^2(x-x')^2}{4a^2}\right]\\
             &\times& \int_{\mathbb R}\rho(x-y,x'-y)
             \exp\left[-\frac{a^2y^2}{v^2}\right]\,{\rm d}y.
\label{finalsqueezrho}\end{eqnarray}
Here $v$ is the norm of the vector $\gamma_1$ in
Eq.~(\ref{xi_i}) and
$u^2=\sum_{i=1}^{k-1} u_i^2$, where $\{u_i\}$ are the coefficients of the
expansion $\alpha_j=\sum_{i=1}^{k-1} u_i \beta_{ji}$, $j=2,\dots,k$.
The parameters $u$ and $v$ quantify the degree to which the secret
state has been degraded for a given $a$ by encoding and decoding.
Perfect replication corresponds to $u=0$ and $v=0$, which is in
general unachievable. The degradation is symmetric under the exchange
of $u\longleftrightarrow v$.

Eq.~(\ref{finalsqueezrho}) shows that the effect of using finite
squeezing for the encoding procedure is twofold. First, the Gaussian
factor in front of the integral in Eq.~(\ref{finalsqueezrho})
supresses off-diagonal elements of the density operator for $x-x' \gg
2a/u$ implying decoherence.  Second, the density operator element of
the replicated secret is a convolution of the original density
operator with a Gaussian.  The larger $a$ is, the more accurately is the
secret state replicated; in the limit $a\to\infty$, it is perfectly
replicated.

\begin{figure}[h]
\includegraphics[angle=270,width=8cm]{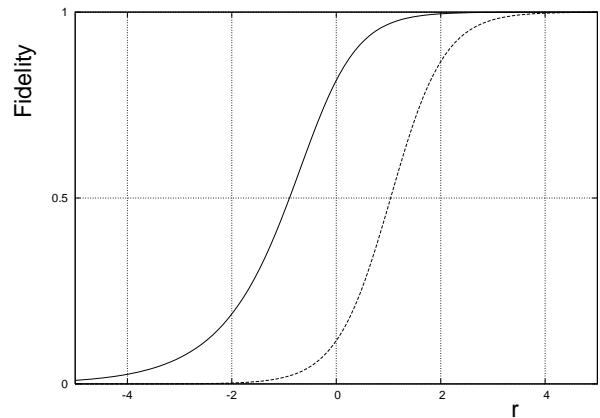}
\caption{The fidelity ${\cal F}$ versus the squeezing parameter $r=\ln a$
  for an arbitrary coherent state as the secret. Two cases are
  presented: (1) $u=0.5$ and $v=1$ (solid line) and (2) $u=3$ and $v=5$
  (dashed line).}
\label{fidelity}\end{figure}

The {\em replication fidelity\/} of the system can be characterized by
evaluating ${\cal F}=\langle\psi|\hat\rho'|\psi\rangle$ for some
standard secret state $|\psi\rangle$.  For an arbitrary coherent state as the
secret, the fidelity is given by the function
\begin{equation}
{\cal F}=[1+(u^2+v^2)/2a^2+u^2v^2/4a^4]^{-1/2},
\label{F}\end{equation}
The dependence of ${\cal F}$ on $r=\ln a$ for some particular values
of $u$ and $v$ can be seen in Fig.~\ref{fidelity}.  The fidelity tends
to unity for large squeezing ($a\to\infty,r\to\infty$) and to zero for
large antisqueezing ($a\to0,r\to-\infty$). The fidelity for $r=0$
corresponds to the case when the ancillary states are all vacuum
states.

\section{Efficient replication}
\label{weakening}

In the previous section we have established a replication protocol for
the access structure; here we seek the most efficient protocol, which
minimizes the total number of squeezers (expensive components in an
active interferometer) required.  In the following we show that by a
suitable choice of a particular disentangling transformation, it is
possible to reduce the total number of squeezers required to no more
than two.

Let $\xi_i\to\zeta_i$ denote the orthogonal projection of $\xi_i\in
\mathbb{F}^n$ to the subspace $\mathbb{X}\oplus \mathbb{Y}\subset
\mathbb{F}^n$ so that 
\begin{eqnarray}
\xi_i({\bf x}) &=& \zeta_i({\bf x}) + \sum_j \gamma_{ij} z_j
\,,\nonumber\\
\zeta_i({\bf x}) &=& \alpha_i x + \sum_j\beta_{ij} y_j \,.
\label{eq:decomp1}
\end{eqnarray}

\medskip \noindent{\em Claim:}
A transformation $g_i\to \xi_i=\alpha_i+\beta_i+\gamma_i$, with
$\alpha_i\in \mathbb{X}$, $\beta_i\in\mathbb{Y}$, and
$\gamma_i\in\mathbb{Z}$,  which leaves the coordinates $\xi_i=g_i$
unchanged for $i=k+1,\ldots, n$ and is such that 
\begin{eqnarray}
  &\alpha_1 =1\,,\quad \beta_1 =0\,,& \nonumber\\
&{\rm span}(\zeta_2,\ldots , \zeta_k)= {\rm
span}(\zeta_{k+1},\ldots,\zeta_n)\,,& \label{eq:effxi}
\end{eqnarray}
disentangles the secret state for sufficiently large values of the
parameter $a$.
\medskip

To prove this claim, we show, by a change of variables that, for
sufficiently large values of $a$, the state
\begin{eqnarray}
|\Psi_\xi\rangle &=& |\det \xi |^{1/2}\int {\rm d}x\,\psi(x)\prod_{i=2}^k
\frac{1}{(\pi)^{1/2}}\int {\rm d}y_i\, {\rm d}z_i \nonumber\\
&&\times\, \exp 
\Big[-\frac{1}{2a^2}  y_i^2 -\frac{a^2}{2}
z_i^2\Big] \nonumber\\
&&\times\, |x+\gamma_1({\bf x})\rangle \otimes |\zeta_2({\bf x})
+\gamma_2({\bf x})\rangle \nonumber\\
&&\quad\otimes \cdots  \otimes 
|\zeta_n({\bf x}) +\gamma_n({\bf x})\rangle  \label{eq:claim}
\end{eqnarray}
defined by the transformation $g_i\to \xi_i$, is expressible in the
form
\begin{equation}
|\Psi_\xi\rangle = |\psi\rangle \otimes |\Phi\rangle
\end{equation}
with
\begin{eqnarray}
|\Phi\rangle &=& |\det \xi |^{1/2}
 \prod_{i=2}^k
\frac{1}{(\pi)^{1/2}}\int {\rm d}y_i\, {\rm d}z_i \nonumber\\
&&\times\, \exp 
\Big[-\frac{1}{2a^2}  y_i^2 -\frac{a^2}{2}
z_i^2\Big] \\
&&\times\,  |\beta_2({\bf x}) +\gamma'_2({\bf x})\rangle
\otimes \cdots  \otimes 
|\beta_n({\bf x}) +\gamma'_n({\bf x})\rangle \,.\nonumber
\end{eqnarray}  

This result is achieved by first changing the variable $x$ to $x-\sum_j
\gamma_{1j} z_j$ and noting that, if $a$ is sufficiently large, then
$\psi(x-\sum_j\gamma_{1j} z_j)\approx\psi(x)$ for all values of
$\sum_j\gamma_{1j}z_j$ for which $\exp\big[-\frac{a^2}{2}
\sum_iz_i^2\Big]$ is non-negligible.
This shows that
\begin{eqnarray}
|\Psi_\xi\rangle &=& |\det \xi |^{1/2}\int {\rm d}x\,\psi(x)\prod_{i=2}^k
\frac{1}{(\pi)^{1/2}}\int {\rm d}y_i\, {\rm d}z_i \nonumber\\
&&\times\, \exp 
\Big[-\frac{1}{2a^2}  y_i^2 -\frac{a^2}{2}
z_i^2\Big] \nonumber\\
&&\times\, |x\rangle \otimes |\zeta_2({\bf x})
+\gamma'_2({\bf x})\rangle \nonumber\\
&&\quad\otimes \cdots  \otimes 
|\zeta_n({\bf x}) +\gamma'_n({\bf x})\rangle \,.
\end{eqnarray}
Next observe that, since the vectors $\{\zeta_{k+1},\ldots,\zeta_n\}$
are linear combinations of the vectors $\{ \zeta_{2},\ldots,\zeta_k\}$,
the change of variables given by the projection
$\zeta_i =\alpha_i + \beta_i \to \beta_i$, for $i=2,\ldots ,k$, results in
the corresponding projections $\zeta_i \to \beta_i$ for $i=k+1,\ldots, n$.
Now, if $\beta^{ij}$ is defined such that
\begin{equation}
\sum_j \beta^{ij}\beta_{jk} = \delta_{ik} \,,
\end{equation}
then the projection $\zeta_i\to\beta_i$, for
$i=2,\ldots ,k$, corresponds to the coordinate
transformation
$y_i\to y_i- (\sum_j\beta^{ij}\alpha_j)\, x$.
Thus, if $a$ is sufficiently large that
$\exp \big[-\frac{1}{2a^2}  (y_i- (\sum_j\beta^{ij}\alpha_j)\,
x)^2\big]\approx \exp \big[-\frac{1}{2a^2}y_i^2\big] $ for all values of
$x$ for which $\psi(x)$ is non-negligible, we obtain
Eq.~(\ref{eq:claim}).

Now, let the vectors $g_i$ defining the encoded state $|\Psi_g\rangle$
(\ref{psi_g}) by the linear transformation (\ref{eq:entangle}) have
decomposition, parallel to that given by Eq.~(\ref{eq:decomp1}),
\begin{eqnarray}  g_i = \kappa_i + \lambda_i \,,&\quad& i =
1,\ldots , k \,, \nonumber\\
g_i = \xi_i = \zeta_i + \gamma_i \,,&\quad& i =
k+1,\ldots , n \,,
\end{eqnarray}
with $\kappa_i\in\mathbb{X}\oplus\mathbb{Y}$ and
$\lambda_i\in\mathbb{Z}$, respectively.
And let $T$ denote a transformation 
\begin{equation}
  g_i \to \xi_i = \sum_{j=1}^k T_{ij} g_j \,, \quad i = 1, \ldots ,k
\end{equation}
such that the vectors
\begin{equation}
 \zeta_i = \sum_{j=1}^k T_{ij} \kappa_j \,, \quad i = 1, \ldots , k
\end{equation}
satisfy the disentanglement criteria (\ref{eq:effxi}).

The condition that the vectors $\zeta_2,\dots,\zeta_k$ span the same
subspace of $\mathbb{X}\oplus\mathbb{Y}$ as do
$\zeta_{k+1},\dots,\zeta_n$ can be satisfied by requiring that both
sets are orthogonal to a common vector
$v\in\mathbb{X}\oplus\mathbb{Y}$.  Thus, if
$v\in\mathbb{X}\oplus\mathbb{Y}$ is a vector defined such that
\begin{equation}
  v\cdot \zeta_{i} = 0 \, \quad i>k \,,
\end{equation}
the transformation $T$ is required to satisfy the equation
\begin{equation}
  v\cdot \zeta_{i} =\sum_{j=1}^k T_{ij}\,v\cdot \kappa_j=0 \,,\quad
\forall\, i=2,\ldots k\,.
\label{cond}\end{equation}
To satisfy the first condition of Eq.~(\ref{eq:effxi}), $T$ should
also be such that
\begin{equation}
  \zeta_1 = \sum_{j=1}^k T_{1j} \kappa_j = f_1 \label{cond1}
\end{equation}
so that $\zeta_1({\bf x}) = x$.

Eq.~(\ref{cond1}) implies that the first row of the matrix $T$ is the
row vector $a=(a_1, a_2, \ldots, a_k)$ whose components are the
coefficients in the expansion $f_1 =\sum_{j=1}^k a_{j}\kappa_j$, i.e.,
$T_{1j} = a_j$. The remaining rows can be defined as a set of
orthogonal row vectors $\{T_i;i=2,\ldots, k\}$, all of which are
orthogonal to the unit row vector $W_1$ whose components are given by
\begin{equation}
W_{1j}= \frac{v\cdot \kappa_j}{\sqrt{\sum_{i=1}^k
(v\cdot \kappa_i)^2} } \,.
\end{equation}
The orthogonality of the vectors $\{T_i;i>1\}$ to $W_1$ then ensures
that $\sum_{j=1}^k T_{ij}W_{1j}=0$ for $i>1$ and that the condition
(\ref{cond}) is satisfied.  The norms of the orthogonal vectors $\{T_i
; i>1\}$ are arbitrary and can be chosen to minimize the cost of the
transformation.  We find (cf.\ following section) that it is
convenient to choose all but one of these vectors (e.g., the vector
$T_2$) to be normalized to unity. Denoting the norm of the vector
$T_2$ by $\gamma$, we then have
\begin{eqnarray}\nonumber
   T_{1j}& =& a_j\\
   T_{2j}& =& \gamma W_{2j} \\\nonumber
   T_{ij}& =& W_{ij}\,,\quad i>2 \,, 
\end{eqnarray}
where $W_{ij}$ is an orthogonal matrix.

As remarked above, an orthogonal transformation of the collaborating
players' states can be achieved with passive elements.  However, the
replacement of the first row of $W$ by the vector $a$, in forming the
matrix $T$, means that the resulting transformation involves squeezing
operations and hence a need for active elements.  As we now show,
the transformation defined by $T$ can be achieved with just two
squeezers.

Choose the vector $W_2$ to lie in the span of the vectors $a$ and
$W_1$.  It then follows that $a$ is expandable as $a=\alpha W_1+\beta
W_2$ and
\begin{equation}
T=\left(\begin{array}{cc|ccc}
\alpha & \beta & 0 & \dots & 0 \\ 0     &  \gamma  & 0 & \dots & 0 \\\hline
 0     &  0    &  & & \\\vdots&\vdots &  & I & \\ 0     &  0    &  & & 
\end{array}\right)W \equiv VW\,,
\label{T:decomposition}
\end{equation}
with a free parameter $\gamma\not=0$.  This parameter can be adjusted,
according to the criteria outlined in Sec.~\ref{total} to minimize the
demands on the squeezing resources.  The ${\rm GL}(k,\mathbb{R})$
matrix $V$ can now be factored as $V=XV_dY$, with $X$ and $Y$
orthogonal matrices and
\begin{equation}
 V_d=\text{diag}(v_1,v_2,1,1,\dots,1).
\end{equation}
The
complete transformation $T$ then assumes the simple form
\begin{equation}
  T=VW=XV_d YW=XV_d Z,
\label{decomp}\end{equation}
with both $X$ and $Z$ orthogonal matrices.

\begin{figure}[h]
\epsfig{file=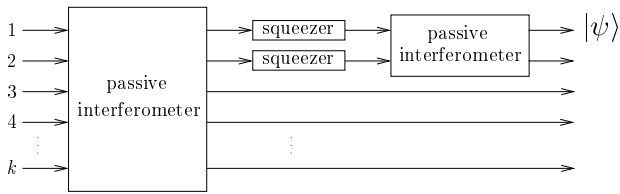, height=1in}  
\caption{The general scheme of an interferometer used by the players to decode
  the secret state. The passive $k$-port interferometer is followed by two
  independent single-mode squeezers, and the last step is a passive two-mode
  interferometer that yields the secret at one output port.}
\label{interferometer}\end{figure}

The disentangling transformation represented by the matrix $T$ is now
achieved by a sequence of three transformations: the first
transformation, represented by the orthogonal matrix $Z$, is achieved
by a passive interferometer consisting of only beam splitters and
phase shifters; the transformation represented by the diagonal matrix
$V_d$ is given by single-mode ${\rm Sp}(1,{\mathbb R})$ squeezers on
the first two modes, with squeezing parameters $r_1=\ln v_1$ and
$r_2=\ln v_2$; finally, the transformation corresponding to the matrix
$X$ is given by a two-mode beam splitter (see
Fig.~\ref{interferometer}).  Hence the number of active optical
elements (squeezers) is reduced to two.

\section{Total amount of squeezing}
\label{total}

It is of interest not only to consider the number of active optical
elements necessary for the replication part of QSS, but also the total
amount of squeezing $R$.  It is natural to define this quantity as the
sum of magnitudes of squeezing parameters corresponding to the two
squeezers, i.e., 
\begin{equation} 
R=|r_1|+|r_2|=|\ln v_1|+|\ln v_2|,
\label{R}\end{equation}
which can be minimized by a judicious choice of $\gamma$ in
Eq.~(\ref{T:decomposition}).

We can express $R$ as $R=\frac12( |\ln\lambda_1|+|\ln\lambda_2|)$,
where $\lambda_{1,2}$ are the eigenvalues of the symmetric matrix
$V'\tilde V'$ with
$V'=\left(\begin{array}{cc}\alpha&\beta\\0&\gamma\end{array}\right)$,
and $\tilde V'$ the transpose of $V'$.  A simple calculation shows
that the eigenvalues are
\begin{equation}
  \lambda_{1,2}=\frac 1 2
  \left[\alpha^2+\beta^2+\gamma^2\pm
  \sqrt{(\alpha^2+\beta^2+\gamma^2)^2-4\alpha^2\beta^2}\right]\,.
\end{equation}

Depending on $\gamma$, the total amount of squeezing $R$ is either (i)
$R=\frac12|\ln(\lambda_1\lambda_2)|$ (if both $\ln\lambda_1$ and
$\ln\lambda_2$ have the same sign) or (ii)
$R=\frac12|\ln(\lambda_1/\lambda_2)|$ (if $\ln\lambda_1$ and
$\ln\lambda_2$ have different signs).  We seek $\gamma$ that minimizes
$R$, which can occur for either case (i) or (ii), so both must be
checked. We define the quantity 
$\kappa\equiv(1-\alpha^2-\beta^2)/(1-\alpha^2)$ and have:
\begin{itemize}
\item[(i)] The minimum value of $R(\gamma)$ is $R_{\rm
    min}=|\ln(\kappa\alpha)|$ and occurs for $\gamma_{0}=\sqrt\kappa$
  in the following situations:
\begin{itemize}
\item[{}] $\alpha^2+\beta^2<1$ and
$\alpha^2+\beta^2<\kappa
$
\item[{}]  $\alpha^2+\beta^2>1+\beta^2/\alpha^2$ and
$\alpha^2+\beta^2>\kappa$
\end{itemize}
\item[(ii)] The minimum value of $R(\gamma)$ is $R_{\rm
    min}=\ln[(\sqrt{\alpha^2+\beta^2}+|\beta|)/|\alpha|]$ and occurs
  for $\gamma_{0}=\sqrt{\alpha^2+\beta^2}$ in the following
  situations:
\begin{itemize}
\item[{}] $1\le\alpha^2+\beta^2\le1+\beta^2/\alpha^2$
\item[{}] $\kappa\le\alpha^2+\beta^2\le1$  
\item[{}] $1+\beta^2/\alpha^2\le\alpha^2+\beta^2\le\kappa$
\end{itemize}
\end{itemize}

The strategy for a collaborating group of players to minimize the
squeezing resources for the replication of the secret state is the
following: for given $\alpha$ and $\beta$, the players calculate the
value of $\kappa$ and decide which of the two cases (i) or (ii)
occurs. Then they find the value $\gamma_{0}$ and construct the matrix
$T$ in Eq.~(\ref{T:decomposition}) and from this, the corresponding
active interferometer that contains only two squeezers with a minimum
total amount of squeezing equal to $R_{\rm min}$.

\section{Conclusion} 
\label{conclusion}

We have shown that the replication procedure in optical
continuous-variable quantum secret sharing can be achieved with a
small number (at most two) of squeezing elements for any authorized
group of players.  In particular, we have demonstrated this for the QSS
threshold schemes. We have quantified the total amount of squeezing
defined as the sum of absolute values of the single-mode squeezing
parameters, and found its minimum value analytically.  We have also
seen that in the realistic situation when the dealer has only finite
squeezing resources available, the density operator of the replicated
secret becomes a Gaussian convolution of the original secret state.

\acknowledgments We would like to thank Martin Rowe for assistance
with calculations of total squeezing.  This project has been supported
by a Macquarie University Research Grant and by an Australian Research
Council Large Grant.

\end{document}